\documentclass[12pt]{article}
\usepackage{graphics}
\begin{document}

\title{\bf{Theory of Manganites Exhibiting Colossal Magnetoresistance}}
\author{T. V. Ramakrishnan$^{\ast \,\dag}$,
\and H. R. Krishnamurthy$^{\ast \,\dag}$,\\
\and S. R. Hassan$^{\ast \,\dag}$ and  \and G. Venketeswara
Pai$^{\ast \,\ddag}$ }

\maketitle
{\raggedright{ {\it{ $^{\ast}$Centre for Condensed Matter Theory,
Department of Physics, Indian Institute of Science,
Bangalore-560012, India\\
$^{\dag}$Condensed Matter Theory Unit, Jawaharlal Nehru Centre for
Advanced  Scientific Research, Jakkur P.O , Bangalore-560064,
India\\
$^{\ddag}$The Abdus Salam International Centre for Theoretical
Physics, 11 Strada Costiera, Trieste 34014, Italy}}} }

\begin{abstract}
The electronic properties of many transition metal oxide systems
require new ideas concerning the behaviour of electrons in solids
for their explanation. A recent example, subsequent to that of
cuprate superconductors, is of rare earth manganites doped with
alkaline earths, namely $Re_{1-x}A_x MnO_3$, which exhibit
colossal magnetoresistance, metal insulator transition and many
other poorly understood phenomena. Here we show that the strong
Jahn Teller coupling between the twofold degenerate ($d_{x^2
-y^2}$ and $d_{3z^2 -r^2}$) $e_g$ orbitals  of $Mn$ and lattice
modes of vibration (of the oxygen octahedra surrounding the $Mn$
ions) dynamically reorganizes the former into a set of states
(which we label $\ell$) which are localized with large local
lattice distortion and exponentially small intersite overlap, and
another set (labelled $b$) which form a broad band. This hitherto
unsuspected but microscopically inevitable $coexistence$ of
radically different $\ell$ and $b$ states, and their relative
energies  and occupation as influenced by doping $x$, temperature
$T$, local Coulomb repulsion $U$ etc., underlies the unique
effects seen in manganites. We present results from strong
correlation calculations using the dynamical mean-field theory
which accord with a variety of observations in the orbital liquid
regime (say, for $0.2\stackrel{<}\sim x \stackrel{<}\sim 0.5$).We
outline extensions to include intersite $\ell$ coherence and
spatial correlations/long range order.
\end{abstract}

\section{Introduction}

Solid state oxides containing transition metal ions with
unfilled $d$ shell electrons have been at the centre of attention
in condensed matter and materials physics for the last two decades
because of the variety and novelty of electronic phenomena in them
and the possibility of new applications.  The $d$ electrons in
solids are not as extended as $s\,,p$ electrons or as localized as
$f$ electrons.  Their motion in the system is highly constrained
by the large local (on site) $d-d$ repulsion or Mott-Hubbard
correlation $U$. The central question is how this strong
correlation, and  other factors such as orbital degeneracy and
electron lattice coupling, lead to electronic behaviour
qualitatively different from that of conventional solids which are
successfully described as degenerate Fermi liquids with well
defined interacting electronic quasi-particles.

\begin{figure}
\resizebox{\textwidth}{!} {\includegraphics[0in,0in][8in,
8in]{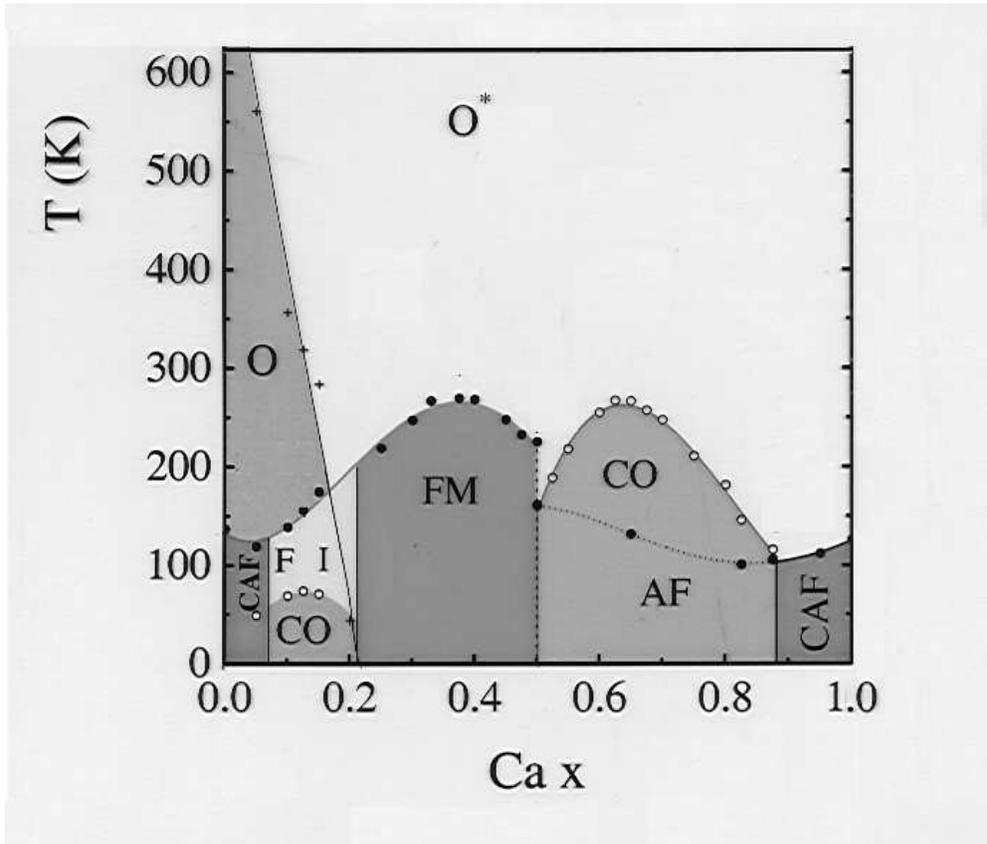}}
\caption{The phase diagram of
$La_{1-x}Ca_xMnO_3$ in the doping $x$ and temperature $T$ plane
(adapted from Ref. \cite{cheong}). Various kinds of
anti-ferromagnetic insulator(AF), paramagnetic insulator(PI),
ferromagnetic insulator(FI), ferromagnetic metal(FM) and
charge/orbitally ordered insulator(CO/O $O^{*}$) are the phases
shown. O and $O^{*}$ are Orthorombic(Jahn-Teller distorted) and
Orthorombic(Octahedron rotated) structural phases.} \label{phdiag}
\end{figure}

While the best known examples of these are the high $T_c$ cuprate
superconductors \cite{Anderson}, during the last decade another
group of oxides, namely the manganites $Re_{1-x}A_x MnO_3$ (where
$Re=La, Pr, Nd$ etc. and $A= Ca, Sr, Ba$ etc.) has  become a focus
of major activity. The initial interest was sparked by the
discovery \cite{Jin} that their electrical resistivity changes
enormously with the application of a magnetic field, the change
(colossal magnetoresistance or CMR) being two or more orders of
magnitude larger than the normal cyclotron orbital effect
characterized by the dimensionless parameter $(\omega_c \tau )^2$.
Subsequent work has shown a bewildering variety of phases, phase
transitions and phenomena \cite{YTokura,Salamon} depending on the
doping $x$, temperature $T$, and ionic species $Re$ and $A$ as
well as external perturbations.  An example is the phase diagram
of $La_{1-x} Ca_x MnO_3$ (Fig. \ref{phdiag}) which shows an
insulating, $Mn-O$ bond (Jahn-Teller) distorted but structurally
ordered phase for small $x$, the transition of this at low $T$
from a ferromagnetic insulator to a ferromagnetic metal at
$x\simeq 0.2$, and thence to a charge ordered insulating phase for
$x \stackrel{>}\sim 0.5$. For $0.2\stackrel{<}\sim x
\stackrel{<}\sim 0.5$ this becomes a paramagnetic insulator above
a $T_c \stackrel{>}\sim 250 \, K$, with CMR near $T_c$. The phase
diagram varies considerably with the ionic species. For example,
$La_{1-x}Sr_x MnO_3$ has a paramagnetic metallic phase for $0.175
\stackrel{<}\sim x \stackrel{<}\sim 0.5$ and shows no
charge/orbital order , while $Pr_{1-x}Ca_x MnO_3$ has no metallic
phases (even ferromagnetic)\cite{YTokura,Salamon}. Two other
general characteristics are the following. First, physical
properties are extremely sensitive to small perturbations;
examples being the CMR itself, the unusually large strain and ion
size effects \cite{Littlwood,Hwang,Attfield}, the melting of
charge/orbital ordering for anomalously small magnetic fields and
a metal to insulator transition induced by the electronically
benign substitution of $O^{16}$ by $O^{18}$ \cite{Babushkina}.
Secondly, over a wide range of $x$ and $T$, two very different
types of regions, one insulating and locally lattice distorted and
the other metallic and undistorted,  coexist \cite{Dagotto}. The
regions can be static \cite{Uehara,Louca} or dynamic
\cite{Louca,meneghini,Heffner}; their size can vary from $100 \,
A^{\circ}$ \cite{Fath} to $3000 \, A^{\circ}$ \cite{Uehara}. All
these observations suggest that metallic and insulating phases are
always very close in free energy.

These phenomena cannot be described satisfactorily in terms of
well known models for electrons  in solids.  For example, one
might expect generally that doping $LaMnO_3$ (a Mott insulator)
with $Ca$ introduces mobile holes in the $Mn-d$ band so that the
system ought to be metallic at least in the high symmetry
paramagnetic phase; but it is not till $x\simeq 0.20$. The
observed properties of manganites need to be understood in terms
of the active degrees of freedom which are believed to be the
twofold degenerate  $e_g$ electrons and the $t_{2g}$ core-spins of
$Mn$, and Jahn-Teller(JT) optical phonon modes of the oxygen
octahedra and the interactions between them. There are three
strong interactions present, namely the large on site $d-d$
repulsion $U \simeq 5 \, eV$ \cite{Satpthy,DDSharma} amongst the
$e_g$ electrons, strong JT mode - $e_g$ electron coupling $g$
which splits the doubly degenerate $e_g$ level by about
$2E_{JT}\,\simeq 1 \, eV$ \cite{MillisRoyalSc,Kanamori}, and the
large ferromagnetic Hund's rule coupling $J_H$ between $e_g$ and
$t_{2g}$ spins ($\simeq$ 1 to 2 $eV$ \cite{Satpthy,DDSharma}). For
comparison, the $e_{g}$ electron  kinetic energy scale or band
width $W \equiv 2D_o$ is $\simeq 2 \, eV$ \cite{Satpthy,DDSharma}.
The presence of a large JT coupling is clearly indicated by the
two `phase' coexistence\cite{Dagotto} (with one 'phase' being
lattice distorted and insulating, and the other metallic and
undistorted) as well as by the sensitivity of properties to
lattice strain or local disorder, and giant isotope
effect\cite{Babushkina,Zhao} alluded to above. Indeed, large JT
distortions, while long ranged in $LaMnO_3$, persist locally well
into the metallic regime\cite{Louca,meneghini} ($x
\stackrel\sim{<} 0.3$, and $T \stackrel\sim{>} 77 \, K$), where
they are short ranged. Understanding the observed consequences of
these interactions is one of the major challenges in the physics
of strongly correlated electrons.

Most existing theories
\cite{Furkawa,MillisMuller,ACGreen,HRoder,Dagotto} for manganites
neglect one or the other of these strong interactions, make
further approximations, and are inadequate qualitatively and
quantitatively. The earliest theoretical approaches
\cite{Furkawa}, commonly referred to as double exchange theories,
consider solely the Hund's rule exchange $J_H$. However, only a
metallic state is possible in this case. A theory due to Millis,
Mueller and Shraiman \cite{MillisMuller} additionally includes the
effect of the coupling $g$, but treats the local JT lattice
distortion classically, as annealed static disorder, and neglects
$U$. A polaronic insulating phase also occurs in that case for
large enough $g$, but the predicted results do not resemble
experiments; for example at $x \neq 0$, one finds only metal-metal
or insulator-insulator Curie transitions, unlike the commonly
observed metal-insulator transition. The magnetoresistance is not
colossal and there is no isotope effect. Dagotto and coworkers
(see eg. Ref. \cite{Dagotto}) have done extensive numerical
simulations of several models, but on rather small lattices, and
seen lots of instances of "phase separation". The early
simulations explored the competition between (double exchange
induced) ferromagnetism and anti-ferromagnetic (super)exchange,
amplification and generation of small scale phase co-existence by
disorder, and identified the two kinds of magnetic domains with
metallic and insulating regions. Later simulations explored
similar issues including static JT couplings. Based on these, they
suggest\cite{Dagotto} that transport in manganites should be
pictured in terms of a random resistor network arising from
tunnelling between misaligned ferromagnetic metallic domains
across insulating regions, and that the CMR arises from enhanced
tunnelling due to field induced alignment of the magnetic domains.
There are many other models designed to address specific effects,
but no theoretical ideas which explain the novel general features
of manganites cohesively.

We propose here a new approach to the physics of manganites which
includes all the three strong interactions, the double degeneracy
of $e_g$ orbitals as well as the quantum dynamics of phonons.  It
leads to a simple physical description from which the properties
of manganites follow naturally. Calculations of ground state and
transport properties at finite temperatures are also presented. We
believe that the ideas are of wide relevance to a large number of
solid state and molecular systems with similar ingredients.

\section{\bf Coexisting polaronic and band states}

Our ideas are principally based on two facts; namely the two fold
degeneracy of the $e_g$ orbital and the strong (degeneracy
lifting) JT interaction characterized by the large dimensionless
ratio $(E_{JT}/\hbar \omega_0)$, ($\simeq 10$ for manganites, see
ref.\cite{MillisRoyalSc})  where $\hbar \omega_0$ is the JT
optical phonon energy $\stackrel{<}\sim 0.075 \, eV$ \cite{Raman}.
This coupling leads (as shown in greater detail below) to a large
local JT distortion $Q_0$  being associated with one linear
combination of the orbitals (JT polaron, labelled $\ell$ by us)
when it occupied by a single electron. If the JT modes are not
approximated as  static displacements \cite{MillisMuller} but are
treated quantum-dynamically, the inter-site hopping of the JT
polaron is reduced by the exponential Huang Rhys \cite{Rhys}
factor $\eta \equiv \exp \{ -(E_{JT}/{2\hbar \omega_{0}}) \}
\simeq (1/200)$ for $E_{JT}=0.6 \, eV$ and $\hbar \omega_{0}\simeq
0.06 \, eV$ . This is the {\em{antiadiabatic limit}} overlap
between the initial and final JT phonon wave functions (centred
around $Q_{0}$ and $0$) when the $\ell$ electron moves in or out
of a site. The corresponding effective $\ell$-bandwidth, $W^{*}
\equiv k_{B}T^{*} \simeq W \eta \simeq  k_{B}(125 \, K)$, is thus
very small, so the $\ell$ states are easily localized by any
disorder present (eg. in cation site energy). They can hence be
regarded, to a first approximation (which may be inaccurate for
$T<T^{*}$ because of inter-site coherence, see Section 8), as non
dispersive localized levels. The $\ell$ polaron is probably close
to this limit over a wide range of $x$ and $T$, where large local
$JT$ distortion without long range order is seen
\cite{Louca,meneghini}.

Since the $e_g$ orbital at each site is doubly degenerate
initially, there is another orthogonal set of states which we
label $b$, which have their largest amplitudes at the fraction $x$
of hole sites where the polaron is not present; their occupancy on
the polaron site costs a large extra energy $\bar U =
(U+2E_{JT})$. The (bare) hopping $\bar t$ amongst these $b$ states
is not reduced and they form a broad band (of bare width $2D_o
\simeq 2$ eV ) whose properties are strongly affected, eg. their
effective bandwidth $2D$ is renormalized to smaller values, by the
other two strong interactions present in the system, namely the
repulsive scattering from the $\ell$ polarons ($\bar U$) and the
coupling to the $t_{2g}$ spins $(J_H)$ depending on $x$ and $T$.
Roughly, $2D$ increases with $x$ as well as with $T^{-1}$ and $H$,
because the inhibition of $b$ hopping due to large $\bar U$ is
reduced when there are more hole-sites, and that due to large
$J_H$ is reduced when the $t_{2g}$ spin order is enhanced.

We believe that the unique feature of manganites is this necessary
coexistence of antiadiabatic, JT distorted, localized $(\ell)$
states and adiabatic, undistorted, broad band $(b)$ states,
arising from a {\em{spontaneous reorganization of the doubly
degenerate $e_g$ states due to the large JT coupling and quantum
phonon dynamics}}. This dynamically generated coexistence is
qualitatively different from that of localized $f$ and extended
$s,p,d$ electrons in rare earth solids, which has an atomic
origin. We show below that for a wide range of $x$ and $T$, both
sets of states are partially occupied. This, we believe, underlies
the ubiquitous two `phase' coexistence observed \cite{Dagotto}.
The high sensitivity of the physical properties of manganites to
small perturbations arises because the $(\ell)$ states, being
localized and lattice distorted, are strongly influenced by local
perturbations, which then affects the delicate relative stability
of $\ell$ and $b$ states.

Finally, (again as discussed in more detail below,) the existence
of the localized polaronic $\ell$ states {\em{in the presence of
large $U$ and $J_H$}} give rise to a {\em{new, major, doping
dependent ferromagnetic nearest neighbour exchange coupling}}
$J_F$ between the $t_{2g}$ core spins. This comes about due to
{\em{virtual, fast (adiabatic)}} hopping processes of the $\ell$
electrons to neighbouring empty sites and back, leading to a $J_F$
roughly of order $x (1-x) {\bar t}^2/( 2E_{JT} S^2)$, the
intermediate state energy due to the unrelaxed lattice distortion
being $2E_{JT}$. We get $J_F \simeq 2 meV$ for $x=0.3$. Our
calculations suggest that this {\em{virtual, correlated double
exchange due to the localized $\ell$ electrons}}, and not
conventional double exchange due to mobile $e_g$ electrons as
hitherto believed, is the dominant source of ferromagnetism and of
the ferromagnetic transition temperature $T_c$ in the hole doped
manganites.

\section{A new model Hamiltonian for manganites in
the strong electron lattice JT coupling regime}

Based on the above ideas, and for the purposes of making
quantitative calculations and predictions, we have
proposed\cite{tvr-prl,gvp-epl} a new model Hamiltonian for
manganites given by
\begin{eqnarray}
H_{\ell b} &=& \sum_{i,\sigma} -(E_{JT}+\mu) \ell^+_{i\sigma}
\ell_{i\sigma} -\mu \sum_{i,\sigma} b^+_{i\sigma} b_{i\sigma}\\
\nonumber &-& \bar t \,\sum_{\langle ij\rangle,\sigma}
b^+_{i\sigma} b_{j\sigma} + U\sum_{i,\sigma } n_{\ell i\sigma}
n_{bi\sigma} + H_s \label{eqhlb}
\end{eqnarray}
Here $\ell^+_{i\sigma}$ creates the JT polaronic state of energy
$-E_{JT}$ and spin $\sigma$, localized at site $i$
($\ell_{i\sigma}$ is the corresponding destruction operator). The
broad band electron, created by the operator $b^+_{i\sigma}$, has
mean energy zero, and nearest neighbour effective hopping
amplitude $\bar t$. The two repel each other on site with energy
$U$. The common chemical potential $\mu$ is determined by the
constraint that the filling, i.e., the average number of $e_g$
electrons per site must be determined by the doping $x$ according
to
\begin{equation}
N^{-1}\sum_{i} \langle n_i \rangle \equiv N^{-1}\sum_{i}
\sum_{\sigma} ( \langle n_{\ell i\sigma}\rangle + \langle
n_{bi\sigma}\rangle) = (1-x). \label{eqnconstr}
\end{equation}

The term $H_s$ in Eq. \ref{eqhlb} models the spin dependent
interactions, and is given by
\begin{eqnarray}
H_s &=& -J_H \sum_i {\vec{s}}_i \cdot \vec{S}_i - J_F \sum_{<ij>}
\vec{S}_i \cdot\vec{S}_j - \mu_B \sum \vec{S}_i\cdot \vec{h}
\label{eqhs}
\end{eqnarray}
It includes the strong ferromagnetic Hund's rule coupling $J_H$
between the $e_g$ spins ${\vec s}_{i} (\equiv ({\vec s}_{\ell i}
+\vec{s}_{bi}))$ and the (spin $S=\frac{3}{2}$) $t_{2g}$ spins
${\vec S}_{i}$ , the net effective ferromagnetic exchange coupling
$J_F$ between these $t_{2g}$ spins (from the mechanism alluded to
above, and discussed in more detail below), and the interaction of
the latter with an external magnetic field $\vec{H}$
\cite{com-lJH}. For simplicity, in what follows we further
approximate the $t_{2g}$ spins as classical fixed length spins
whose directions fluctuate, i.e. write $\vec S_i = S
{\hat{\Omega}}_i$ where ${\hat{\Omega}}_i$ is a unit vector.

We now discuss how the above model Hamiltonian can be motivated
from the microscopic Hamiltonian for the manganites in the limit
of strong JT interaction. A reasonably realistic microscopic model
Hamiltonian or energy operator $H$ of $d$ electrons ($e_g$ and
$t_{2g}$) in the manganites has three types of contributions,
namely those involving only the $e_g$ electrons ($H_e$), the
coupling of these electrons with the JT lattice modes as well as
the energy of these lattice modes themselves (${H}_{e\ell}$), and
the part $H_s$ which involves the $e_g$ and $t_{2g}$ spins. Thus

\begin{equation}
H= H_e + H_{el} + H_s \label{eqhmicro}
\end{equation}

where
\begin{eqnarray}
H_e &=& \sum_{i,\sigma}(\epsilon_i -\mu) \tilde{n}_{i\sigma} +
\sum_{<ij>}\,\tilde{a}^+_{i\sigma}\tilde{t}_{ij}\,
\tilde{a}^+_{j\sigma} \\ \nonumber &+& U
\sum_{i,\sigma\sigma\prime}
(\tilde{n}_{i\sigma}\tilde{n}_{i\sigma'}-\tilde{n}_{i\sigma})
\label{eqhe}
\end{eqnarray}

\begin{eqnarray}
H_{el} \equiv \sum_i (H^i_{JT} + H^i_l) &=& g \sum_{i,\sigma}
\left(\tilde{a}^+_{i\sigma} \tau^z \tilde{a}_{i\sigma} Q_{3i} +
\tilde{a}^+_{i\sigma}\tau^x \tilde{a}_{i\sigma} Q_{2i}\right)\\
\nonumber &+& \frac{K}{2} \sum_{i}(Q^2_{2i} + Q^2_{3i}) +
\frac{1}{2M} \sum_i \left(p^2_{2i} + p^2_{3i}\right) \nonumber
\label{eqhel}
\end{eqnarray}
Here the operators $\tilde{a}^+_{i\sigma},\tilde{a}_{i\sigma}$
respectively add and remove an electron at site $i$, in the spin
state $\sigma$ which can take two values, and in the two $e_g$
orbital states with labels $\alpha = 1 $ corresponding to $d_{3z^2
-r^2}$ and $\alpha = 2 $ corresponding to $d_{x^2-y^2}$; i.e.,
$\tilde{a}^+_{i\sigma}$ is a short hand for $(a^+_{i 1
\sigma}\,,a^+_{i 2 \sigma})$. The number operator
$\tilde{n}_{i\sigma}$ is the usual sum $\sum_{\alpha}
a^+_{i\alpha\sigma}\, a_{i\alpha\sigma} \equiv
\tilde{a}^+_{i\sigma} \tilde{a}_{i\sigma}$.  The $e_g$ electron
has site energy $\epsilon_i$, equal to 0 for a clean system.  The
(anisotropic and orbital dependent) nearest neighbour hopping
{\em{matrix}} is $\tilde{t}_{ij}$ and $U$ is the repulsion between
electrons at $i$ in different states. The first set of terms in
$H_{e\ell}$ describes the coupling of $e_g$ electrons with the two
JT lattice modes $Q_{3i}$ and $Q_{2i}$ with strength $g$,
($\tau^a$ being the usual Pauli matrices in orbital space) the
former displacement leading to a splitting and the latter to a
mixing of the two $e_g$ states. The last two terms in $H_{el}$ are
the potential and kinetic energies of the modes neglecting
anharmonic and intersite terms. $H_s$ is the same magnetic
Hamiltonian as in eqn. \ref{eqhs} (except that, in the context of
the microscopic hamiltonian, the $e_g$ spin $\vec{s}_i$ $\equiv
\frac{1}{2} \sum_{\mu \mu '} \tilde{a}^+_{i\mu}
(\vec{\sigma})_{\mu\mu'} \tilde{a}_{i\mu'}$ where $\vec{\sigma}$
is the Pauli spin operator).

In manganites the JT coupling $g$ is large compared to $t/Q_0$
(where $Q_0$ is the typical size of the JT distortion), and one
can first concentrate on the single site, coupled electron-phonon
problem. The Schrodinger equation corresponding to
$H^i_{JT}+H^i_l$ can be solved exactly at a single site.
$\vec{Q}_i \equiv (Q_{2i}, Q_{3i})$ is like a local
pseudo-magnetic field splitting the pseudo-spin $1/2$ orbital
levels 1 and 2, hence $H^i_{JT}$ has eigenvalues $\pm g Q_i$,
where $Q_i = |\vec{Q}_i|$ is the magnitude of the JT distortion.
The diagonalization of $H^i_{JT}$ can be done explicitly using the
operators
\begin{equation}
b^+_{i} \equiv cos (\theta_i /2) {a^+}_{i1} + sin (\theta_i /2)
{a^+}_{i2}
\end{equation}
and
\begin{equation}
{\tilde{\ell}}^+_{i} \equiv -sin (\theta_i /2) {a^\dag}_{i1} + cos
(\theta_i /2) {a^+}_{i2},
\end{equation}
where $\theta_i \equiv tan^{-1} (Q_{2i}/ Q_{3i})$, i.e., it
determines the orientation of the JT distortion. When an electron
is present at site $i$ in the $\tilde{\ell}$ state, the effective
lattice potential energy at site $i$ in $H^i_{JT}+H^i_l$ has the
form
\begin{equation}
V^-_i = (K Q^2_i/2 - g Q_i)
\end{equation}
This has a minimum at $Q_0 = (g/K)$ (the JT polaronic distortion)
with an energy lowering of $E_{JT}= (g^2 /2K)\sim$ .5 eV
independent of $\theta_i$ \cite{Sturge}. The JT polaronic $\ell$
state that we referred to above is this $\tilde {\ell}$ state {\it
together with the JT distortion}. When an electron is present in
the other `anti' JT , $b$ state, the effective lattice potential
energy is
\begin{equation}
V^+_i = (K Q^2_i/2 + g Q_i)
\end{equation}
with a minimum at zero displacement so that no distortion is
associated with it.

Thus, when an $\ell$ polaron is present at a site, low energy
lattice configurations around the deep minimum of $V_i^-$ involve
a shift of $Q_i$  by an amount $Q_0$ at each site. This is
formally achieved by a Lang-Firsov transformation \cite{LangF}
${\hat {\eta}}_i \equiv exp \left\{-(i p_i Q_0/\hbar)\right\}$
where $p_i$ is the radial momentum operator conjugate to $Q_i$. In
other words, formally, one has ${\ell}^+_{i} \equiv {\hat
{\eta}}_i {\tilde{\ell}}^+_{i} $. The Huang-Rhys reduction
factor\cite{Rhys} $ \eta \equiv exp \left\{ -(E_{JT}/ 2\hbar
\omega_0) \right \}$ we have alluded to earlier, which multiplies
the hopping involving the $\ell$ states, is just the ground state
expectation value of ${\hat {\eta}}_i$, corresponding to the
anti-adiabatic limit overlap of the distorted phonon wavefunction
at a site with the undistorted phonon wavefunction. (We ignore
fluctuations on the basis that ${\bar t} \eta \ll \hbar\omega_0$
\cite{PaiTVR}). The coherent intersite hopping of the $\ell$
electrons is thus $t^* = {\bar t}\eta \simeq 10 \, K$, very small,
and can be neglected to a first approximation. The `anti' Jahn
Teller state $b$ has no associated lattice distortion, so its
hopping amplitude $t_{ij}$ is not reduced, but depends on the
angles $\theta_i$ and $\theta_j$ at sites $i$ and $j$. We assume a
nearest neighbour hopping amplitude $\bar{t}_{ij}= \bar t$
averaged over angles $\theta_i$ either statistically (and
classically) or by quantum fluctuations . Such an {\em{orbital
liquid}} approximation is reasonable for the phases with no long
range orbital order, i.e., for $0.2 \stackrel {<}{\sim} x
\stackrel {<}{\sim} 0.5$ in most manganites, but poor for those
other values of $x$ for which one has strong orbital correlations
or long range orbital order.

The localized polaronic $\ell$ and the broad band $b$ states
coexist in manganites since the conditions for anti-adiabaticity
of the former, namely $\bar t \eta \ll \hbar \omega_0$ and the
adiabaticity of the latter, namely  $\bar t \gg \hbar \omega_0$,
are {\em{both}} fulfilled. The $b$ electrons are repelled strongly
by the $\ell$ electrons, with energy $\bar U = U + 2 E_{JT}$, and
hence will have the largest amplitudes at sites where the $\ell$
electrons are not present, i.e. at the hole sites. They hop
quickly (time scale $\hbar/ \bar t \ll 1/\omega_0$) among such
sites, avoiding strongly the sites with static $\ell$ electrons,
where they hence have small amplitudes\cite{com-percol}.

Finally we discuss the result alluded to above namely that the
existence of localized, JT distorted $\ell$ states in the presence
of large $J_H$ and large $U$ can give rise to a new mechanism of
ferromagnetic exchange between the $t_{2g}$ core spins, which we
refer to as "virtual double exchange".

Suppose that a localized $\ell$ electron is present at a site, and
that site is distorted. The $\ell$ electron can take part in fast
(adiabatic) virtual hopping processes to neighbouring sites, i.e.,
leaving the local lattice distortion unrelaxed, by paying an
energy cost of $2E_{JT}$ in the intermediate state. For large $U$
this can happen only if the neighboring state is empty, and for
large $J_H$ only if the $t_{2g}$ spins on the two sites are
parallel. (Otherwise the energy of the intermediate state will
increase by $U$ and $J_H$ respectively.) Clearly, from second
order perturbation theory this process will give rise to a
{\em{new, $\ell$ occupancy dependent, ferromagnetic exchange
coupling}} between the $t_{2g}$ core spins of the form
\begin{equation}
\left( \frac {{\bar t}^2}  {2E_{JT}} \right ) \frac{1}{2} ( {\hat
\Omega}_i \cdot {\hat \Omega}_j + 1 ) [n_{\ell i}(1-n_{j}) +
n_{\ell j}(1-n_{i})]
\end{equation}
(where we have again ignored the dependence on the angles
$\theta_i$ etc). The $\frac{1}{2}({\hat \Omega}_i \cdot {\hat
\Omega}_j + 1 )$ factor comes from large $J_H$, and the occupancy
dependent terms from large $\bar U$. Within the homogeneous
orbital liquid approximation we are using in this paper, this
translates to an effective ferromagnetic interaction proportional
to $x {\bar n}_\ell \simeq x(1-x)$ as stated earlier.

The normal super-exchange coupling between $t_{2g}$ core spins on
neighboring sites is of order ${\bar t}^2/U$ , and its sign
depends on the nearest neighbour orbital correlations
\cite{Kh-Sa}. For example, in $LaMnO_3$ it is anti-ferromagnetic
along the c axis, and ferromagnetic in plane. Conventional double
exchange can, in principle, explain metallic ferromagnetism in the
manganites, but would be unable to account for ferromagnetism in
their insulating phases in the absence of orbital correlations.
The above mechanism yields a ferromagnetic coupling which is
clearly much larger than and dominates the super-exchange for all
intermediate doping, operates even in the insulating states and
even in the orbital liquid phase, and as we discuss later, makes
the dominant contribution to the ferromagnetic $T_c$ even in the
metallic case.

Putting all this together, expressing the on site coulomb and
exchange interactions in the $\ell$ and $b$ basis, and including
the new exchange mechanism, leads us to the suggestion that in the
large $(E_{JT}/\hbar\omega_o$) limit, and for describing low
energy electronic properties the microscopic model Hamiltonian in
Eq. \ref{eqhmicro} can be approximated by the much simpler, two
fermion species, Hamiltonian $H_{\ell b}$ given in Eq.
\ref{eqhlb}. The relatively small $g \mu_B \vec{H} \cdot
\vec{s}_i$ term  and other coulomb repulsion terms (between $\ell$
and $b$ electrons of different spins)  have been neglected in
$H_{\ell b} $ because of the large $J_H$ ($J_H \gg \bar t$) limit
of relevance here (whence the spin of the $\ell$ and $b$ electrons
is forced to be parallel to the $t_{2g}$ spin), and we have
relabelled the effective repulsion between the two types of
electrons as $U$.

\section{DMFT wth polaronic and band states }

Apart from the constraint equation Eq. \ref{eqnconstr} and the
spin dependent $H_s$, $H_{\ell b}$ is essentially the well known
Falicov-Kimball  model (FKM)\cite{fk} for non hybridizing $f$
electrons in a broad band metal with correlations. Indeed at
$T=0$, in the ferromagnetic phase and for large $J_H$, the spin
degrees of freedom are completely frozen, whence it reduces to the
(spin-less) FKM. More generally, it describes the dynamics of the
$b$ electrons moving in an {\em{annealed disordered}} background
of immobile $\ell$ electrons and $t_{2g}$ spins in the presence of
strong on site repulsion $U$ and Hund's coupling $J_H$ , the
annealed disorder distribution being {\em{thermodynamically and
self-consistently determined}}. This strong correlation problem
cannot be solved in general, but can be solved within the
framework of dynamical mean field theory (DMFT) \cite{Georges}
which is exact at $d=\infty$, and is quite accurate for three
dimensions. We have carried out these DMFT calculations for
thermodynamic properties (the occupancies $\bar{n}_\ell$,
$\bar{n}_b$, the magnetization $m \equiv \langle
\vec{S}_i\rangle$, the specific heat, magnetic susceptibility,
etc.), spectral behaviour (eg. the $b$ band self-energy,
propagator and density of states(DOS)) and transport (resistivity
and magnetoresistance). While the specific results we discuss
below come from such calculations, we believe that many of the
physical mechanisms that underlie the results have a larger sphere
of validity.

The properties of $H_{\ell b}$ are determined largely  by the
dynamics of the $b$ electrons, eg., their propagator or Green's
function $G_{ij}(\omega)$, as affected by the (annealed
disordered) distribution of $ n_{\ell i}$ and $ {\hat{\Omega}}_i$
due to strong $U$ and $J_H$. In the DMFT framework, the $b$
electron self energy $\sum_{ij}(\omega)$ due to these interactions
is site local, i.e. $\sum_{ij}(\omega) = \delta_{ij}
\sum(\omega)$, and is determined self consistently. The general
approach is well known \cite{Georges}.

Essentially, the problem reduces to that of one site, with its
local degrees of freedom and interactions, immersed in an
effective medium (comprising the other sites). We make the
simplest reasonable approximation corresponding to a homogeneous
annealed system, namely that $\langle n_{\ell i\sigma}\rangle
\equiv \bar{n}_{\ell \sigma} $ , $\langle n_{b i\sigma}\rangle
\equiv \bar{n}_{b \sigma}$ and $\langle \hat{\Omega}_i \rangle
\equiv \vec m = m \hat{z}$ are the same at every site $i$, and
that there are no correlations between these quantities at
different sites\cite{com-mms}. Consistent with this, we
approximate the ferromagnetic exchange interactions in $H_s$ in
terms of a homogeneous molecular field in the standard way and
replace
\begin{equation}
H_s = -J_H S \sum_i {\vec s}_i \cdot {\hat {\Omega}}_{i} - ({\vec
h} + {\bar J}_F \vec m ) \cdot \sum_i {\hat {\Omega}}_{i}
\label{eqhsmofld}
\end{equation}
Here $\vec{h} \equiv g \mu_B S \vec{H}$ , ${\bar J}_F \equiv 2zJ_F
S^2$ where $z$ is the co-ordination number. The quantity $m $,
proportional to the magnetization, and $\bar{n}_{\ell \sigma} $,
the average occupancy of the polaronic states, are also to be
determined self-consistently.

Thus, in the present context, the effective medium is a
homogeneous electron bath with which the local $b$ electron
hybridizes,  the molecular field $\bar{J}_F m$ in Eq.
\ref{eqhsmofld}, and the local distributions of $ n_{\ell \sigma}$
and $\hat {\Omega} $ . The on-site $b$ electron propagator due to
the medium, but without the local interactions, is ${\cal G}$
where the (Matsubara) frequency variable and the spin label are
suppressed. The single site problem can be solved exactly, just as
in the Falicov-Kimball case \cite{Georges},  to determine (1) the
(homogeneous) annealed distributions $P(n_{\ell})$ and $P({\hat
\Omega})$ as functionals of ${\cal G}$, $m$ and the model
parameters and (2) the local $b$ electron propagator $G$ as a
functional of ${\cal G}$ ,$m$, the model parameters and these
annealed distributions. There are two other formal relations among
$G$, $\cal G$ and $\Sigma$, namely $(G^{-1}={\cal G}^{-1}-\Sigma)
$ and $ G=\sum_{\vec {k}}[({G_{\vec k}^{0}}^{-1}-\Sigma)^{-1}]$.
By iterating these equations to self-consistency , one can
determine $\Sigma$, $G$, ${\cal G}$ , $P(n_{\ell})$ and  $P({\hat
\Omega})$ (and hence $\bar{n}_{\ell \sigma}$, $\bar{n}_{b \sigma}$
and $m$) explicitly for any chosen values of the model parameters
$U$, $\epsilon_{\ell}$, $\bar t$ and $\bar{J}_F$ at a fixed $T$
and $\mu$ (the latter being chosen such that $\bar{n}_{\ell}+
\bar{n}_{b}=(1-x)$). Thermal, spectral and transport properties
can be calculated in a straightforward way \cite{Georges} from the
$b$ electron propagator. For example, the current-current
correlation function (the Kubo formula) which determines the
electrical (and optical) conductivity, can be expressed entirely
in terms of $G_{ij}(\omega)$, vertex corrections being negligible.

The resulting calculations, while straightforward, still involve
extensive numerics. The calculations whose results are described
below have been done in the large $J_H$ limit, i.e. assuming that
the spin of the $\ell$ and $b$ electrons are forced to be aligned
along $t_{2g}$ spin ${\vec \Omega}$, whence there is considerable
simplification \cite{MillisMuller,com-mms}. We have done such
calculations both for a realistic model $\bar{t}$ and for a model
semicircular density of states. The two results are not very
different, and here we mostly discuss results for the latter case,
which has the advantage that some exact analytical results can be
obtained in the $U=\infty$, $T = 0$ limit, and even the finite
$U$, non-zero $T$ calculations are much simpler. The details are
described in refs. \cite{gvp-epl,tvr-prl,has-the} and other papers
we will publish elsewhere.

\section{Metal insulator transitions}

To start with, we show how the ideas above lead to a simple
physical understanding of insulator metal transitions (IMT) in
manganites, based on the variation of the occupancies and relative
energies of the $\ell$ and $b$ states with $U, x, T$ etc.

\begin{figure}
\resizebox{\textwidth}{!} {\includegraphics[0in,0in][8in,
8in]{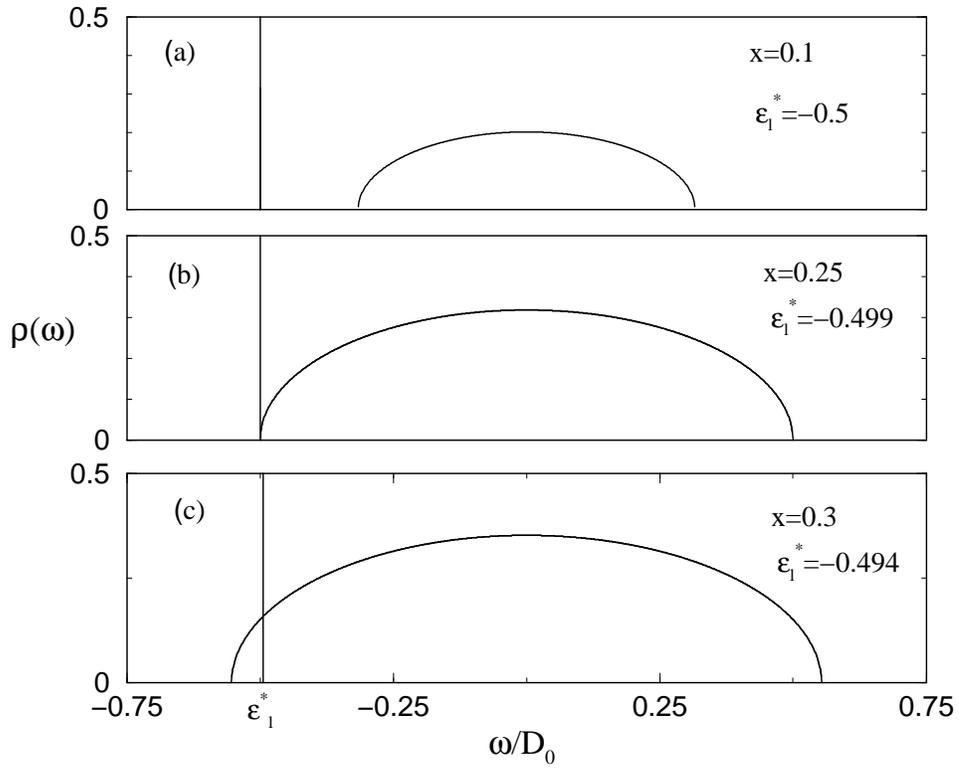}} \caption{Evolution of DOS of the $b$ band
for various doping values of $x=$ 0.1, 0.25, 0.3 and $T = 0$ with
parameters $E_{JT} =  0.5 \, eV $, $D_o = 1 \, eV$. The effective
$\ell$ level, labelled as $\varepsilon_{\ell}^*$ is also
indicated. } \label{figT0DOS}
\end{figure}

Fig. \ref{figT0DOS}  shows the density of states for $(E_{JT}/D_o
)= 0.5$ , $D_o=1 \, eV$ , $U= \infty$ (and ${\bar J}_F = 60.2 \,
meV $) for several values of $x$ at $T= 0$.
At $T=0$, because of $J_F$ the  ground state is ferromagnetic, and
the spin degrees of freedom are completely frozen. For large $U$,
the renormalized or effective $b$ bandwidth $D$ self consistently
goes to zero as $x\rightarrow 0$, whence for small $x$ the bottom
of the $b$ band will be above the $\ell$ level (of energy
-$E_{JT}$) and all electrons are in the latter states, as in Fig.
2a. For, the low energy $b$ band states are constrained to reside
mostly on the small fraction $x$ of empty, undistorted sites, and
have a small amplitude at the lattice distorted sites (where
$\ell$ electrons are present) from which they are repelled with
energy $U$ . The system is hence an insulator. This cannot happen
in a pure double exchange model, where ferromagnetism and
metallicity go together. As $x$ increases, so does $D$, and there
is a critical $x_c$ (= 0.25 within DMFT for the parameters used in
Fig.2) at which the $b$ band bottom touches the $\ell$ level
energy as in Fig.2b. For $x>x_c$, the $b$ band bottom goes below
the $\ell$ level, and some of the $\ell$ electrons are transferred
to the $b$ band until all $b$ levels up to the (now renormalized)
$\ell$ level are occupied as in Fig. 2c. The system is then a
ferromagnetic metal, but most of the electrons (0.6 per site out
of the 0.7 present) are in the polaronic $\ell$ state though there
is no long range JT order.

Thus there is an IMT, not at $x=0$ as in a naive doped Mott
insulator picture , but at $x_c \neq 0$. This shift is due to the
strong electron lattice coupling $g$ which makes the JT polaronic
$\ell$ level possible and stabilizes it by an energy $E_{JT}$, and
large $U$, which makes the $b$ electron bandwidth vanish as
$x\rightarrow 0$. For $U \rightarrow \infty$, within DMFT we can
show\cite{gvp-epl} that $D = D_o \sqrt {x} $ whence we obtain
$x_c$ analytically as $ x_c= (E_{JT}/D_o)^2$ where $D_o$ is the
bare half-width of the b-band. The decrease of $x_c$ with
increasing $D_o$  and its numerical value for reasonable choices
of $E_{JT}$ and $D_o$ are in accord with experiments (for example,
as one goes from $La Ca MnO_3$ to $La Sr Mn O_3$ which has a
larger bandwidth). Finite $U$ corrections to $x_c$, which we have
calculated, are small for $U\simeq 5 eV$ which is appropriate to
these systems. A more detailed discussion of these and other zero
temperature results is presented in \cite{gvp-epl}.

\begin{figure}[htbp]
\resizebox{\textwidth}{!} {\includegraphics[0in,0in][10in,
10in]{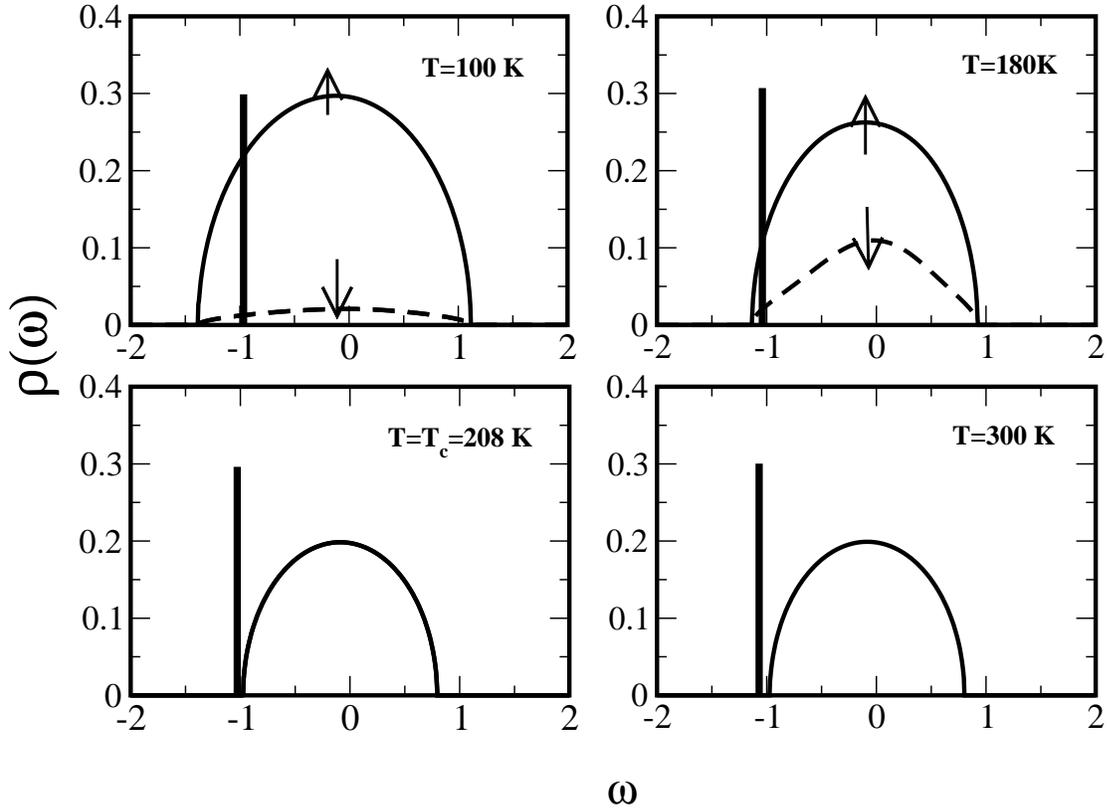}} \caption{Evolution of the spectral
function $\rho_b(\omega)$ for parameter values that correspond to
$Sr$ doping (see text), for $T=$ 100K, 180K, 208K, 300K and $x=$
0.175. Thick lines represent the effective $\ell$ level and up and
down arrow indicate up spin and down spin spectral functions.}
\label{figchap4Sr5}
\end{figure}

Fig.\ref{figchap4Sr5} shows the evolution of the spectral
functions with temperature, for parameter values that correspond
to $Sr$ doping, for a fixed doping of $x=0.175 >x_c$ for this
system, and for the temperature values $T=$100K, 180K, 208K, 300K.
For $x>x_c$, the system is metallic so that both $b$ and $\ell$
electronic states occupied. On increasing the temperature,  the
$t_{2g}$ spin order decreases and the $b$ band narrows again,
since due to double exchange \cite{Hasegawa} the effective hopping
amplitude of the $b$ electrons decreases. The spectral weight for
the down spin polarization also turns on, reflecting the reduction
in the magnetization from its zero temperature value. The sequence
of figures 3a to 3c illustrates this progression. Eventually, both
up and down spin spectral functions become equal, and the $b$ band
bottom crosses the $\ell$ level and moves up as $T$ rises beyond
$T_c$ (= 208 K in the present context, see Fig. 3c) where
ferromagnetism disappears, and the system becomes a paramagnetic
insulator as in Fig. 3d, with $b$ states occupied thermally across
a gap. We thus have a simple picture of the thermal
ferromagnetic-metal to paramagnetic-insulator transition.

\section{Resistivity, CMR and material systematics}

There are very few calculations
\cite{Furkawa,MillisMuller,ACGreen} of the transport properties of
manganites though this is one of their main unusual features.
Within the DMFT for our model, the electrical resistivity $\rho$
can be calculated in a straightforward way\cite{Georges} from the
$b$ electron propagator or Green's function. Fig. 4 shows some of
our results, for $x=$ 0.3 and parameters representative of
$Nd_{1-x}Sr_x MnO_3$ and $La_{1-x}Ca_xMnO_3$ \cite{com-prmt}.
Experimental values are also given for comparison \cite{PDai}.

\begin{figure}
\resizebox{\textwidth}{!} {\includegraphics[0in,0in][8in,
8in]{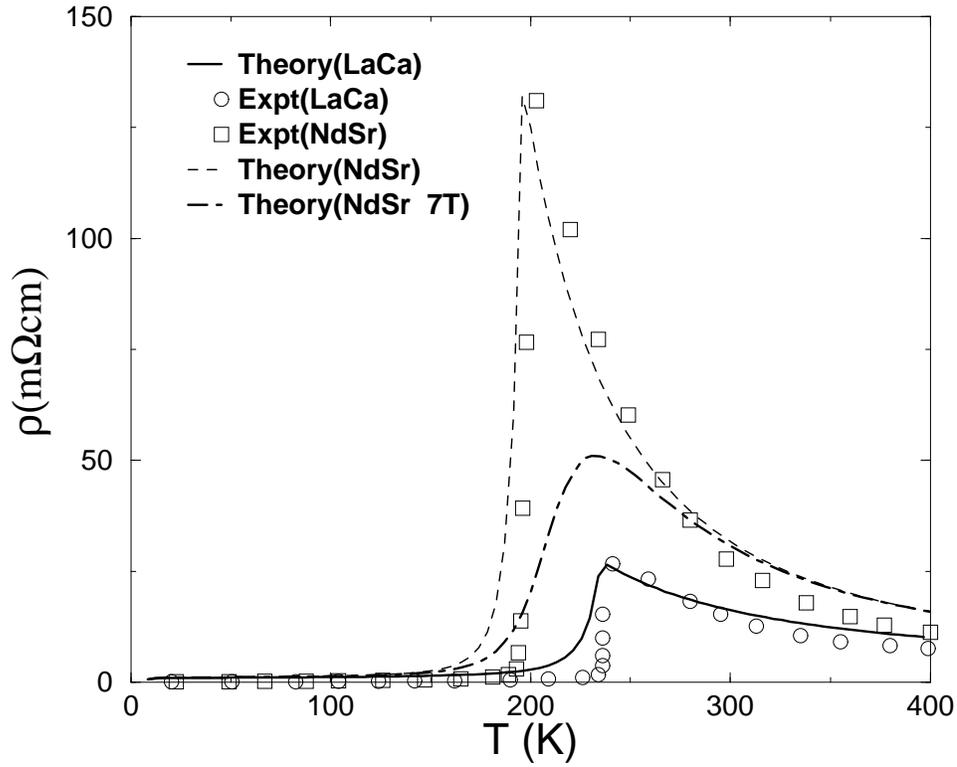}} \caption{Electrical resistivity and CMR in the
coexisting JT polaron - broad band model for $E_{JT} = 0.5 \, eV$,
$U=5 \, eV$ and $x = 0.3$. $D_o$ and ${\bar J}_F$ are chosen so as
to reproduce the experimental\cite{PDai} $T_c$ and ${\rho (T_c)}$
of $Nd_{1-x}Sr_{x}MnO_{3}$ ($D_o=1.05 \, eV, {\bar J}_F = 52.7 \,
meV$) and of $La_{1-x}Ca_{x}MnO_{3}$ ($D_o=1.15 \, eV , {\bar J}_F
= 60.2 \, meV$) . Calculated $\rho(T)$ at $H=7 \, Tesla$ is shown
for $Nd_{0.7}Sr_{0.3}MnO_{3}$}. \label{fig-res-cmr}
\end{figure}

We see that the resistivity of the paramagnetic insulating  state
is fairly well described by the theory. In this phase, the
conducting $b$ band is only thermally occupied. The effective
electrical gap $\simeq 850K$ (the experimental value is $\simeq
1250 K$\cite{PDai}). The positive feedback between the thermally
excited, strongly $T$ dependent number ${\bar n}_{b}(T)$ of $b$
electrons, and the double exchange broadening of the $b$ band as
$T$ decreases, leads to a very rapid decrease and closure of the
electrical gap just below $T_c$ and the consequent rapid decrease
in the resistivity just below $T_c$ as in experiment to a rather
large value $\simeq 2 \, m\Omega cm$. But we note that the
calculated resistivity does not decrease much thereafter down to
$T=0$, unlike experiments, where $\rho(T)$ decreases to residual
values of $\simeq 50 \, \mu \Omega cm$  below about $T = 125 K$.
We believe that the latter is due to inter-site $\ell$ coherence,
neglected here (see later).

We also show in the Fig. \ref{fig-res-cmr} the resistivity in a
field of $7 Tesla$ for $Nd_{1-x}Sr_{x}MnO_3$. We clearly see CMR.
The external magnetic field polarizes the $t_{2g}$ spins, and via
$J_{H}$ increases the $b$ band width which reduces the electrical
gap. CMR arises because resistivity depends exponentially on the
gap. The effect is largest near $T_c$ as the change in the
magnetization due to a given magnetic field is the largest there.
As our mean field approximation for the Curie transition neglects
short range magnetic order above $T_c$, $\rho$ is somewhat
overestimated and the CMR somewhat underestimated.

The properties of manganites vary strongly and characteristically
with the ionic species $Re$ and $A$. The material systematics of
such variation, eg., in the thermal IMT as well as the CMR depend
in our model largely and sensitively on the  ratio $(D_o/E_{JT})$
and somewhat on the $D_o$ and $J_F$. A small increase in
$D_o/E_{JT}$ reduces the high temperature resistivity $\rho(T >
T_c)$ enormously, as is clear in Fig. \ref{fig-res-cmr} where we
have increased $D_o$ from 1.05 to 1.15 $eV$ and ${\bar J}_F$ from
52.7 to 60.2 $meV$ to fit $T_c$ and $\rho (T_c)$ for $La_{0.7}
Ca_{0.3} Mn O_3$. Broadly this is because the density of current
carrying $b$ electrons and hence $\sigma (T) = 1/ \rho (T)$
depends exponentially on $(E_{JT} - D_o)$.  This is also the
general reason for the observed large variation of $\rho(T > T_c)$
and $T_c$ with strain and pressure.

Making the reasonable assumption that $E_{JT}$, a  single
octahedron quantity, is roughly unchanged across materials, but
that ${\bar J}_F$, the effective ferromagnetic exchange, scales as
$D_o^2$, we have calculated properties as a function of $D_o$ for
$x\simeq 0.3$. We find that $T_c$ is typically fairly close to the
(Curie-Weiss) value (${\bar J}_F/3$) (and hence increases as
$D_{o}^2$ as $D_o$ increases), with a slight enhancement due to a
nonzero $\bar n_{b}(T_{c})$ and double exchange.  With increasing
$D_o$, the resistive transition on lowering $T$ through $T_c$
changes from insulator-insulator to insulator-metal, and then to
metal-metal. These trends are seen experimentally \cite{Hwang} as
a function of increasing cation radius $\langle r_A \rangle$,
which is known to roughly track the bandwidth $D_o$. We also find
that the relative CMR, defined as $[R(H)-R]/R(H)$, depends
exponentially on $T_c$ (increasing as $T_c$ decreases). This is
indeed the behaviour extracted from measurements on several
systems \cite{Zettle}. The observed unusually large dependence of
$T_c$ and of the resistivity at $T_c$ on pressure \cite{Neuameir}
in $La_{1-x}Ca_{x}MnO_{3}$ is reproduced if we assume that $D_o$
increases with pressure at a realistic rate of about .01
$eV/Kbar$. A more detailed discussion of these finite temperature
results and material trends is presented in \cite{tvr-prl}.

A key feature of our theory is the result mentioned above that the
dominant contribution to $T_c$ comes from virtual double exchange
of the localized $\ell$ polarons and not from conventional double
exchange, since the mobile carrier density (${\bar n}_b$) is very
small. Interestingly, the precipitous, universal and nearly linear
drop in $T_c$ with local ion size variance reported by
Attfield\cite{Attfield} in a variety of multi-component manganites
is much larger than what is expected from a simple double exchange
picture. In the double exchange mechanism, $T_c$ is proportional
to $t_{ij}$ which depends on the $Mn-O-Mn$ bond angle $\phi_{ij}$,
which is strongly affected by the local ion-sizes, as
$cos(\phi_{ij})$. The variance in $\phi_{ij}$, namely $ < (\delta
\phi_{ij})^2 > ^{1/2} $ is at most 10 degrees, so that one expects
that the resulting reduction in $T_c$ is $\Delta T_c / T_c \sim <
(\delta \phi_{ij})^2 > / 2 \sim 1/60$. The observed reduction is
an order of magnitude larger! We believe that such a large
reduction is however understandable within our theory if we extend
the calculations described above to include the effect of ion size
variance on $E_{JT}$ , and $\ell - b$ hybridization effects as
discussed in section 8.

\section{Other unusual properties }

We now illustrate through three examples how other unusual
properties of manganites can be understood in our approach.
Specifically, we consider the anomalously low carrier density as
inferred from optical conductivity in the metallic phase
\cite{Neff}, the small electronic specific heat \cite{Salamon} and
the electron hole asymmetry \cite{Electrondopedexpt}.

\subsection{Low metallic carrier density}
In the metallic regime $(0.2 \stackrel{<}\sim x \stackrel{<}\sim
0.5)$, one can estimate $n_{eff}$ , the effective density of
carriers, from the the Drude weight i.e. the frequency integral of
the real part of the optical conductivity $\sigma_r (\omega)$. In
a free electron like limit, this integral is $(\pi n e^2 /2m^*)$
where $n$ is the density of carriers, and $m^*$ is the band
optical mass.  Since there is no evidence for enhancement of $m^*$
with respect to the band mass (eg. see Ref.\cite{Salamon}),
results for $La_{1-x} Sr_x Mn O_3$\cite{Neff} lead to $n_{eff}=
0.06$ for $x= 0.3 $ and $T=0$. The expected value (for a doped
Mott insulator) is $(1-x)=0.7$. Although there are several
uncertainties in inferring $n_{eff}$ precisely from
$\sigma_{r}(\omega)$, its reduction by nearly an order of
magnitude, and even more so its observed strong decrease with
temperature on the scale of $T_c$ which is much less than the
(normal) Fermi temperature, are puzzling. But the results are
easily understood in our model, since the low frequency $\sigma
(\omega)$ is due to $b$ electrons. As seen from Fig. (2c) at $T=0$
there are indeed very few electrons in the $b$ band, all near its
bottom, below the $\ell$ (or Fermi) level.  The DMFT calculation
with parameters appropriate for $La_{1-x}Sr_x Mn O_3$ at $x =0.3$,
namely $E_{JT} =  0.45 \, eV \,, D_{o} = 1.28 \, eV \,, {\bar J}_F
= 1150 \, K $ (and consequently $T_c =390 \, K$), leads to $\bar
n_{b}(T=0)=0.12$, and a decrease with temperature which is very
similar to what is seen in experiment.

\subsection{Normal linear electronic specific heat}
The low $T$ electronic specific heat in  manganites (see Ref.
\cite{Salamon} for a review) goes as $\gamma T$ with a $\gamma$
nearly independent of $x$, unlike other doped Mott insulators(eg
Ref. \cite{Salamon}) and has a value expected  for a tight binding
band with bandwidth $2D_o \simeq 2.0 eV$. In polaron theories with
$\bar{n}_{\ell}$ polarons per site $C_{v}={\bar n}_{\ell}k_{B}$ in
the classical gas limit, and $\sim{\bar
n}_{\ell}k_{B}(k_{B}T/D^*)$, where $D^{*}$ is the effective
bandwidth, for $T\ll (D^{*}/k_{B})$. Neither of these is observed
and this is often argued to be evidence against small polarons in
manganites. In our model, although the electrons are mostly JT
polarons, we expect that they will make very little contribution
to the low temperature specific heat. For, being localized  they
will strongly couple to disorder potentials and affected by
coulomb interactions, and we expect that their entropy will be
frozen out at a large temperature scale. We find that $\gamma$
(calculated in the paramagnetic metallic phase to avoid large
magnetic contributions to the specific heat) is about $0.9 \,
(mJ/K^2 mole)$ with the parameters used above for
$La_{0.7}Sr_{0.3}MnO_{3}$. The value for a band of width $2 \, eV$
is about $2.0 \, (mJ/K^2 mole)$. The physical reason for the
smaller $\gamma$ is that the few $b$ electrons which are present
occupy states near the bottom of the $b$ band with energy close to
$-E_{JT}$, and the density of states there is relatively
smaller(Fig. 2c).

\subsection{Electron hole asymmetry}
The  electronic and magnetic properties of hole doped $(x <0.5)$
and "electron doped" ($(x>0.5$) manganites are very different. For
example, the latter are much more metallic than the former in the
paramagnetic phase \cite{Electrondopedexpt} (although the AF or
charge ordered phases can be insulating for other reasons). This
behaviour is unexpected in a one orbital strong coupling
antiadiabatic polaron model since for low $e_g$ carrier density
the dilute small polaron assembly will form an insulator, as also
in a model with only $J_{H}$ , where there is electron hole
symmetry. It is often stated that polaronic effects are not seen
for $e_g$ orbitals in the $e$ doped regime because the JT coupling
becomes weaker as $x\rightarrow1$. This seems unlikely given the
local nature of the JT interaction and the stability of the
($MnO_{6}$) octahedron.

The metallic nature of the  $e$ doped regime which has low  $e_g$
density  is natural in our theory because for large $x$ the
effectively uncorrelated $b$ electrons form a wide band whose
bottom is occupied by the small number of $e_g$ electrons, and the
$\ell$ states are unoccupied, as depicted in Fig. 2d. This fact
also explains why pure band models are fairly successful\cite{Pai}
in describing the magnetic ground states in the the $e$ doped
limit.

\section{Discussion}
In summary, we have presented here a new model of coexisting
localized JT polarons and broad band electrons for manganites,
argued that it arises inevitably in the presence of orbital
degeneracy and strong JT coupling, and shown that it explains a
wide variety of characteristic properties of manganites. In this
concluding section we compare our theory with some of the other
theories, discuss some of the inadequacies of the present model
and extensions to overcome these. We believe that a more general
treatment of the model Eq. \ref{eqhlb} with some extensions can
lead to a complete description of manganites. Two examples, namely
inclusion of spatial correlations and of intersite $\ell$ state
coherence are discussed below.

To begin with, we note that the picture developed here is very
different from that  in other polaronic models, which either
neglect $e_g$ orbital degeneracy \cite{ACGreen,HRoder} or work in
the adiabatic approximation \cite{MillisMuller,Mukul}. In the
former, for example, at high temperatures transport is due to the
activated hopping of localized small polarons. In the latter
(adiabatic) models, the polarons also form a broad band, whence it
is difficult to obtain a paramagnetic insulator to ferromagnetic
metal transition say at $x\simeq 0.3$. Furthermore, small polarons
disappear below $T_c$ even for large $g$, the Drude weight is not
small, and there is no isotope effect, all in disagreement with
experiment. In both, it is argued that small polarons are likely
only at high temperatures, and that small and large polarons
coexist only in a narrow range of $x$ and $T$ determined by the
effective electron phonon coupling and that there are no small
polarons at low $T$.

In our theory, the carriers are broad band electrons thermally
promoted out of localized JT polaronic states. Small JT polarons
(in the anti-adiabatic limit, with negligible bandwidth) and band
states(in the adiabatic limit) necessarily coexist over a wide
range of $x$ $(0.2 \stackrel{\sim}< x \stackrel{\sim}< 0.7$ say)
{\em{and for all}} $T$. In contrast, in single orbital polaron
models, there is a crossover with increasing dimensionless
electron phonon coupling $\lambda_{eff} (x,T) =\left\{g^2/K
\,W_{eff} (x,T)\right\}$ from large to small polarons, so that the
two can occur together only over a narrow range of $x$ and $T$.

{\it An unusual feature of the polaron level in our model is that
it has no prominent thermodynamic or spectroscopic (sharp level)
manifestations.} Firstly as pointed out above, it does not lead to
a large electronic specific heat as we expect that their entropy
will be frozen out at energy scales determined by coulomb and
impurity interactions not explicitly included in our model. The
remaining linear specific heat for the $b$ band metal, is roughly
of the size seen experimentally\cite{Salamon}. Secondly, no sharp
$\ell$ like spectral feature is to be expected. Rather, the $\ell$
excitation spectrum will be an incoherent continuum, starting from
a weak (weight $\propto \eta \simeq 1/200 $) anti-adiabatic low
energy part building up to adiabatic (Franck-Condon) higher energy
features, as the fast removal of an $\ell$ electron leads to
highly excited lattice states (energy $\sim 2 E_{JT}$). Indeed,
optical conductivity data in manganites do show\cite{Neff} such an
incoherent continuum which is largely independent of temperature,
and we believe that this is their origin.

In our theory presented earlier we made the simple approximation
that the local $JT$ distortion $Q_o$ is independent of $x$ and
$T$, and that its large value is determined entirely by on-site
properties, namely the electron $JT$ phonon coupling $g$ and the
phonon mode force constant $K$.  This is probably valid for high
temperatures not too far below $T_c$ and for
$x\stackrel\sim{<}0.5$. However, JT polaronic effects can become
dynamical and unobservable at low $T$ for two reasons . Firstly,
the local lattice kinetic energy has a term $(1/2)M Q_o^2
\dot{\theta}^2_i$ where $Q_o$ is the magnitude, and $\theta_i$ the
direction or orientation, of the local JT distortion. Hence
$\theta_i$, which also determines the orbital mixing amplitudes of
the $\ell$ and $b$ states, rotates due to quantum fluctuations,
although due to anharmonicity and crystallinity, the rotation
could be hindered and slow. A second, perhaps more important,
source of this dynamical $J-T$ effect is the quantum mechanical
intersite $\ell$ state coherence which makes the $\theta_i$ ill
defined. The energy scale for the latter, $D^* \sim 125 K$, is
consistent with the widely observed
\cite{Salamon,Dagotto,Louca,Heffner} reduction or disappearance of
static or of long time scale local $JT$ distortion in metallic
manganites as their temperature decreases well below 125K.
Finally, intersite $\ell$ coherence, especially hybridization with
energetically degenerate broad band $b$ states in the metallic
regime, will reduce $Q_o$ and broaden the $\ell$ band. The two
effects feed back on each other, and it is in principle possible
that small polarons may weaken and disappear altogether at low
temperatures in the metallic phase rather than merely becoming
dynamic.  We have not explored this possibility here.

Intersite $\ell$ state coherence can be included in our model by
adding to Eq.(1) an $\ell - b$ hybridization  term $\sum_{\langle
ij \rangle ,\sigma} \left[ t_1 ({\ell^+}_{i\sigma} b_{j\sigma} +
{b^+}_{j\sigma}\ell_{i\sigma}) + t_2 \ell_{i\sigma}^+
\,\ell_{j\sigma} \right]$ where $t_1 \sim {\bar t} \eta \sim D^*$,
and $t_2 \sim {\bar t} \eta^2$ . We expect that its inclusion, and
the consequent development of long range intersite $\ell$ state
coherence at low temperatures will, in addition, lead to the
observed smooth decrease \cite{Zhao} in the electrical resistivity
of clean metallic manganites from about $1 \, m \Omega cm$ just
below $T_c$ to a small value, $\simeq 50-100 \, \mu \Omega cm$, as
$T\rightarrow 0$ past a characteristic crossover temperature of
about 100K, of the same scale as $D^*$. In the DMFT calculations
presented above, the residual resistivity at $T=0$ due to random
scattering from the incoherent $\ell$ sites is too large, $\sim 1
\, m \Omega cm$ (Fig.3). If the $\ell$ states become coherent, the
$b$ electron scattering and the consequent resistivity would then
vanish at $T=0$, and is nonzero only if static disorder is
present. This can lead to a metallic state with a small residual
resistivity or to an Anderson localized insulating state depending
on the amount of disorder.

We believe that the giant isotope effect observed in manganites
\cite{Zhao} is another dramatic consequence of the $\ell - b$
hybridization  scale $t_1$ and its exponential dependence on the
inverse square root of the isotopic mass. For it can add, via
(conventional) double exchange, to the ferromagnetic $T_c$ an
amount roughly given by $\Delta T_c \simeq \alpha \bar{n}_\ell
(D^*/k_B) \simeq (87 \alpha) \, K$ for $\bar{n}_\ell =0.7$ where
$\alpha$ is less than one. Because of the Huang Rhys factor in
$D^*$, which depends exponentially on square root of the isotopic
mass, we find that $\Delta T_c (O^{16}) \simeq (1.33)\, \Delta T_c
(O^{18})$ so that for $\alpha \simeq (1/2)$ the difference
$T_c(O^{16}) - T_c(O^{18}) \sim 15 \, K $, close to the observed
value\cite{Zhao}.

In addition to the neglect of intersite coherence effects
discussed above, we have also neglected in our theory spatial
correlations between the local electron densities $n_{\ell i}$ and
$n_{bi}$, as well as between the local angles $\theta_{i}$. This
may be adequate for the phases which correspond to {\em{homogenous
orbital liquids}}. However, many phenomena in manganites such as
short range order, long range charge/orbital order (or
{\em{orbital solidification}}) , various types of
antiferromagnetism \cite{YTokura,Salamon} and mesoscale structures
\cite{Hwang} depend  on spatial correlations.  To treat these we
need to add to eq. (1) a number of more complicated longer range
coulomb, anharmonic, steric, elastic, magneto-elastic, etc.
interactions that couple spin, orbital and lattice degrees of
freedom to each other, and to strain, ion size mismatch, disorder,
etc. as appropriate. A self consistent determination of long or
short range order in $\delta_i \equiv (\bar{n}_{\ell
i}-\bar{n}_{bi}) $ (a new `internal' variable)\cite{com-cd} ,
$\theta_i$ and ${\vec{S}}_i$, and their effects on the $b$
electron dynamics for different $x$ and $T$, in the presence of
such interactions can lead to a complete description of manganites
including the above phenomena, and others such as first order
transitions, two phase co-existence, etc.. We hope to discuss
these elsewhere.

A significant general question raised by our work is that of the
detailed nature of adiabatic to non adiabatic crossover as $g$
increases, and the conditions for the {\em{coexistence of
adiabatic and antiadiabatic states with exponentially separated
dynamical timescales}}.  We have shown that the latter leads to
new phenomena, argued that this happens in manganites because of
the large $g$, and have developed the consequences in a simple
model assuming this separation. The results closely correspond
with a wide variety of observations.  It would be of great
interest to explore such a crossover and time-scale or
energy-scale separation experimentally and theoretically in the
many systems such as organic solids, transition metal oxides and
chemical (molecular) systems, which have degenerate orbitals and
strong symmetry breaking JT couplings.

We would like to acknowledge support from several agencies. HRK
has been supported in part by grant no. 2404-1 of the Indo French
Centre for Promoting Advanced Research, TVR in part by a US-India
project ONR N 00014-97-0988, and SRH and GVP in part by the
Council of Scientific and Industrial Research, India.


\begin{thebibliography}{99}
\bibitem{com-lJH} The relatively small $g \mu_B \vec{H} \cdot \vec{s}_i$
term in $H_s$ and other coulomb repulsion terms in $H_{\ell b} $
have been neglected because of the large $J_H$ ($J_H \gg \bar t$)
limit, of relevance here.
\bibitem{com-percol} In the limit that this amplitude is zero, we have
two classical regions, one with small polarons and another with
broad band states; this is the percolative regime discussed in the
literature (eg. \cite{Dagotto,Uehara}).
\bibitem{com-mms} We note that the approach of Millis, Muller and Shraiman
\cite{MillisMuller} which we have commented on earlier is similar,
except that they set $U=0$ and treat the JT distortions purely
classically.
\bibitem{com-prmt} We choose  $E_{JT} = 0.5 \, eV$
and $U=5.0 \, eV$,  for both the systems. $D_{o}$ is $1.05 \, eV$
and $1.15 \, eV$, and ${\bar{J}}_F$ is $52.7 \, meV$ and $60.2 \,
meV$ respectively. These choices are made such that the observed
$T_c$ as well as $\rho(T_c)$ are reproduced. The electrical
activation energies above $T_c$, $\Delta_{theory}=850 \, K, 550 \,
K$, are in reasonable agreement with $\Delta_{expt}=1250 \, K, 770
\, K$ respectively. Further the steep drop in $\rho$ below $T_c$
is well reproduced.
\bibitem{com-cd} Variations in the local charge density $q_i \equiv \left(\bar{n}_{\ell i}
+\bar{n}_{bi}\right) $ are likely to be small and of short range
because of the large Coulomb energy cost. (A coulomb energy term
needs to be added to Eq. \ref{eqhlb} to describe this).
\end{thebibliography}
\end{document}